\documentstyle[twocolumn,prl,aps]{revtex}
\begin{document} 
\title{
Internal Excitations and Dissipative Damping of Quantum Hall Skyrmions
}
\author{H.A. Fertig$^1$, L. Brey$^2$, R. C\^ot\'e$^3$, and A.H. MacDonald$^4$}

\address{1.
Department of Physics and Astronomy, University of Kentucky,
Lexington, Kentucky 40506-0055.
}

\address{2.
Instituto de Ciencia de Materiales (CSIC),
Universidad Aut\'onoma C-12,
28049 Madrid, Spain.
}

\address{3.
Universit\'e de Sherbrooke, D\'epartement de Physique, 
Sherbrooke, Qu\'ebec, Canada J1K 2R1
}

\address{4.
Department of Physics, Indiana University, Bloomington, IN 47405
}

\address{\mbox{ }}

\address{\parbox{14.5cm}{\rm \mbox{ }\mbox{ }
We propose an intrinsic maximum speed for dissipationless E cross B
drift of Skyrmion quasiparticles in quantum Hall ferromagnets.
When this speed is exceeded, Skyrmions can radiate 
spin-waves by making internal excitations which 
allow total spin to be conserved.  Our proposal  
is illustrated by a time-dependent Hartree-Fock approximation
calculation of the excitation spectrum for a Skyrmion bound
to an impurity. 
}}
\address{\mbox{ }}
\address{\mbox{ }}
\address{\parbox{14.5cm}{\rm PACS numbers: 73.40.Hm, 73.20.Mf, 75.30.Ds }}
\maketitle

Studies of low-energy excitations in quantum Hall systems 
continue to be a rich source of new physical phenomena\cite{prange}.
Recently, it has been found that spin-polarized
incompressible quantum Hall states may support charged quasiparticles
that carry non-trivial spin 
textures\cite{kane,sondhi,hf,moon,sklat,oaknin,hcm} 
(``spin-texture quasiparticles'', or STQ's).
These textured quasiparticles are a natural generalization of
skyrmion states that arise in the context
of the non-linear sigma model\cite{raj},
and have a spin density that may be
characterized by a winding number of precisely unity.
They carry a charge in quantum Hall ferromagnets 
because\cite{kane,sondhi,natoasi} of the commensurability relations 
between magnetic flux and charge density required for incompressibility.
In the ground state, deviation 
from the incompressible filling factor $\nu$ is 
accomplished by introducing STQ's which 
are responsible for a rapid degradation of the system's 
spin polarization\cite{hf}.  This effect has been observed 
near $\nu =1$ in several recent 
experiments\cite{barrett,eisenstein,aifer} 
that directly probe the spin density of the two-dimensional electron gas.
In this letter we address the dynamics of STQ's.  We propose 
an intrinsic limit for the speed of dissipationless  
E cross B drift of STQ's.  When this speed is exceeded, STQ's 
can radiate spin-waves by making internal excitations which
allow total spin to be conserved.  

Since STQ's are finite size solitons,
classical descriptions\cite{raj} suggest they will 
have unusual localized excitations.
Indeed, it follows from  microscopic considerations\cite{hcm} 
that STQ's of both positive and negative charge 
have an internal quantum number $K = 0, 1, \cdots $
which specifies the number of reversed spins in their interiors. 
(The $K=0$ STQ's are the undressed minority spin electron   
and majority spin hole states.) 
It follows that the dependence of the energy of STQ states
on Zeeman energy has the form $\epsilon_K = U_K + K \tilde g$ 
so that their energetic ordering is dependent on the 
Zeeman coupling strength, $\tilde g = g^{*} \mu_B B$.  
Existing experiments\cite{barrett,eisenstein,aifer}
are consistent with $K=3$ for the lowest energy STQ at $\nu=1$ 
under typical experimental circumstances.  

In two-dimensions, a non-interacting charged particle in 
a strong perpendicular magnetic field responds to an external potential
which is smooth on the scale of its cyclotron orbit radius by
drifting the orbit guiding center along an
equipotential\cite{semiclass}.  This dissipationless 
E cross B drift plays an important role in the theory 
of quantum Hall effect transport phenomena and is, in particular,
the basis for the network model\cite{huckestein} of  
localization behavior in a strong magnetic field.  For 
the case of a constant electric field $E$ in the $\hat x$ direction,
the guiding center drifts in the $\hat y$ direction with 
velocity $v_{dr}= c E / B$.  In a quantum treatment, electronic
states within a Landau level are distinquished by 
wavevector $Q_y$ related (up to a gauge dependent constant)
to the $x-$ coordinate of the guiding center by 
$X = \ell^2 Q_y$, where $\ell=\sqrt{\hbar c/eB}$ is the magnetic length.
Dissipative electronic motion 
by a distance $\ell^2 Q_y$ along the electric field 
can take place only if the electronic energy ( $ e E \ell^2 Q_y$ )
and momentum $\hbar Q_y$ change are transfered to long wavelength phonon
excitations of the host semiconductor.  These transitions
are possible only if $v_{dr}$ exceeds semiconductor speed of sound, 
a condition which appears to be connected\cite{vkstreda} with the breakdown 
of the quantum Hall effect at large current densities.
The recent discovery of STQ's implies that the quasiparticles
relevant to quantum Hall transport phenomena near $\nu =1$ 
(in weak disorder systems) are not bare electrons or holes but $K \ne 0$ STQ's.
In the presence of a smooth disorder potential the Landau
levels for STQ's will also exhibit dissipationless 
E cross B drift\cite{c1}.  In the STQ case, however, dissipation 
is possible at sufficiently large drift velocities even 
without phonon emission.  A STQ can 
move along the electric field and conserve energy 
and momentum by emitting a long wavelength spin wave and 
reducing its internal quantum number ($ K \to K-1$) 
to conserve total spin.  This dissipative process is not
available to bare $K=0$ quasiparticles in the absence of 
spin-orbit coupling and is related to   
spin-charge coupling in quantum Hall ferromagnets\cite{kane,sondhi,moon}.
The energy conservation condition
for transitions with momentum transfer $\hbar Q_y$ is:
\begin{equation}
e E \ell^2 Q_y + \epsilon_K - \epsilon_{K-1}= 
\tilde g  + 4 \pi \rho_s \ell^2 Q_y^2  
\label{eq:engcond}
\end{equation} 
The right-hand-side of Eq.(~\ref{eq:engcond}) is the energy 
of a long-wavelength spin-wave\cite{kallin} with wavevector $Q_y$.
This equation has solutions if $v_{dr}$ exceeds 
\begin{equation}
v_{max} \equiv \big( 16 \pi \rho_s \ell^2 (\tilde g + \epsilon_{K-1} 
- \epsilon_K ) \big)^{1/2}.
\label{eq:vmax}
\end{equation} 
$v_{max}$ is the speed limit for dissipationless E cross B drift 
of $K \ne 0$ STQ's\cite{c2}.
This process provides a mechanism for the 
descriptions\cite{stone} of STQ dynamics.

To illustrate the importance of this new dissipative process 
for STQ's we have evaluated the longitudinal linear response
function at zero temperature, using the time-dependent Hartree-Fock
aproximation (TDHFA), for the case of a single STQ bound by a circularly 
symmetric impurity potential.
The circular symmetry of this
situation is compatable with the circular symmetry of 
the quasiparticles and simplifies our calculations.  When the 
impurity potential is smooth E cross B drift of the STQ's takes place
around the circular equipotentials.   For quantum calculations, 
the linear momentum of the above discussion is replaced by an
angular momentum; $Q \to M/R_M$ where $R_M \approx \sqrt{2M} \ell$ is 
the radius of the equipotential corresponding to the $M$'th quantized 
orbital.  The poles of the response functions
occur at the neutral excitation energies of the system. 
{}From the residues of the poles we may learn
whether the excitation has a charge, spin, or mixed character.

As discussed previously\cite{hf}, the Hartree-Fock (HF) approximation
for the ground state of a system with a single (hole-like) STQ 
may be written as
\begin{equation}
|\psi_{-}> = \prod_{m=0}^{M_{max}-1} \alpha^{\dag}_{m+1}
\prod_{m=M_{max}}^{\infty} c^{\dag}_{m+1,\uparrow}|0>
\quad
\label{hfstate}
\end{equation}
where $\alpha^{\dag}_{m+1} =  u_m c_{m,\downarrow}^{\dag} 
+v_m c_{m+1,\uparrow}^{\dag}$ creates states in a linear
combination of spin down states with angular momentum $m$
and spin up states with angular momentum $m+1$ in the 
lowest Landau level.  Note that all single particle states in this Slater
determinant have definite values for the {\it difference} of 
orbital and spin angular momenta. 
The HF approximation to the STQ state is found by minimizing 
Eq. (\ref{hfstate})
with respect to the variational parameters
$u_m$ and $v_m$.  A set of states in the
lowest Landau level that are eigenstates of the mean-field 
HF Hamiltonian and are orthogonal to the $\alpha$
states may be written as $\beta_{m+1}^{\dag}= -v_m c_{m,\downarrow}^{\dag} 
+u_m c_{m+1,\uparrow}^{\dag}$; these represent unoccupied
states of the HF band structure.  In terms of these
states, we can define generalized response functions
of the form
$$
\chi^{\gamma_1,\gamma_2,\gamma_3,\gamma_4}_{n_1n_2}(m,\omega)
=-\int_0^{\infty}dt  e^{i\omega t} \times\quad \quad 
$$
\begin{eqnarray}
<Ta_{n_1,\gamma_1}^{\dag}(t)
a_{n_1+m,\gamma_2}(t) a_{n_2+m,\gamma_3}^{\dag}(0)
a_{n_2,\gamma_4}(0)> 
\label{chidef}
\end{eqnarray}
which have poles whenever $\pm \hbar \omega$ is equal to a
neutral excitation energy.  Here $\gamma_i=\alpha,\beta$, and 
$a_{n,\gamma}=\gamma_n$ is a notation that encompasses
both the occupied and unoccupied HF bands.

In this work, we focus on excitations within the lowest
Landau level so that only the $\alpha$ and $\beta$ bands
need to be retained in the calculation.  Excitations to
higher Landau levels will be discussed in a future
publication\cite{future}.  
To compute $\chi$, we use a variant of the TDHFA
that has been employed with great success 
for electrons in strong magnetic fields 
obeying periodic boundary conditions\cite{cote},
although in our situation, the equations of motion
need to be found for a system with circular symmetry.
Details of our calculational method will be
presented elsewhere\cite{future}.

\begin{figure}
 \vbox to 6.0cm {\vss\hbox to 8cm
 {\hss\
   {\includegraphics{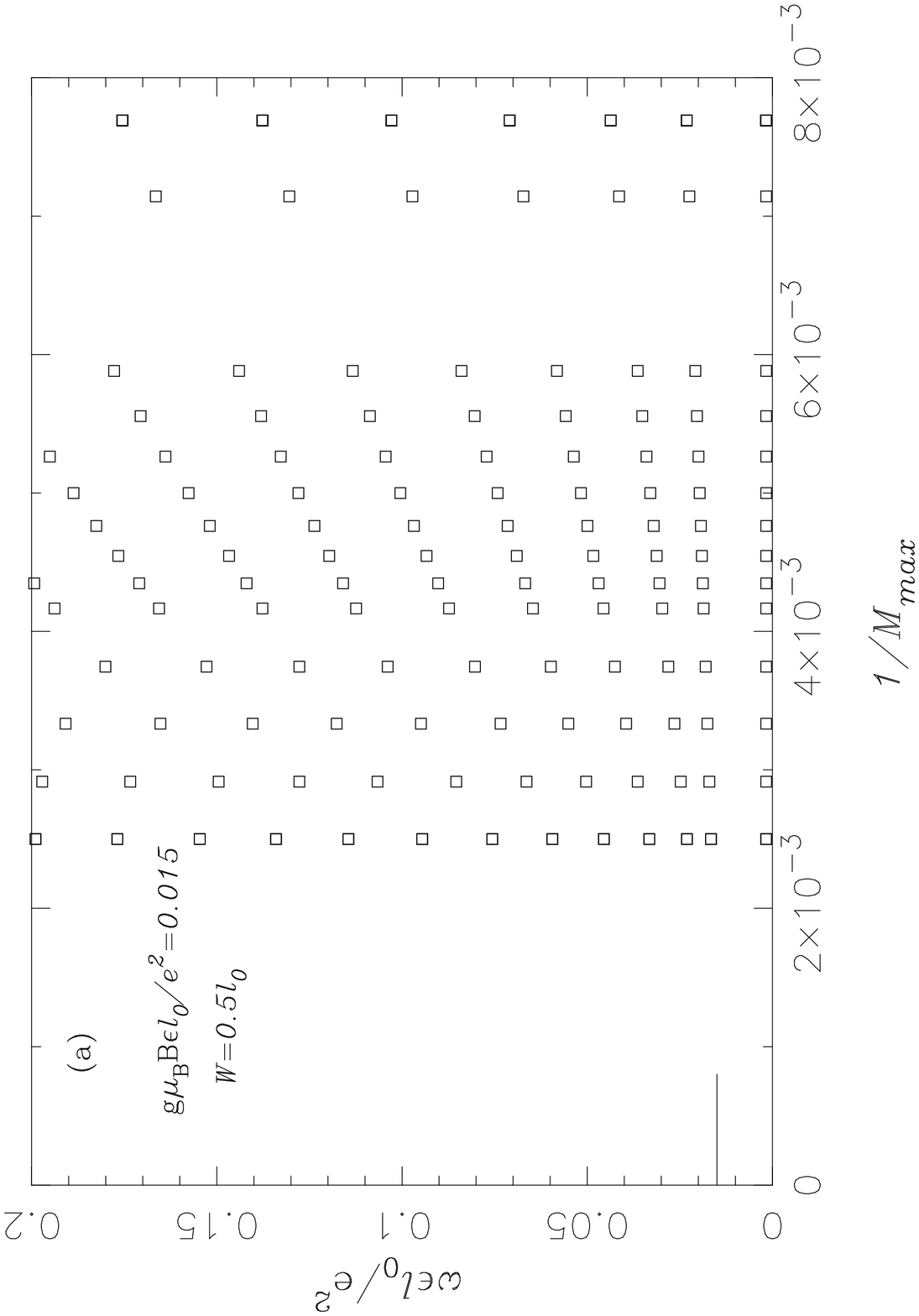}
   }
  \hss}
 }
\vspace{2mm}
 \vbox to 6.0cm {\vss\hbox to 8cm
 {\hss\
 {\includegraphics{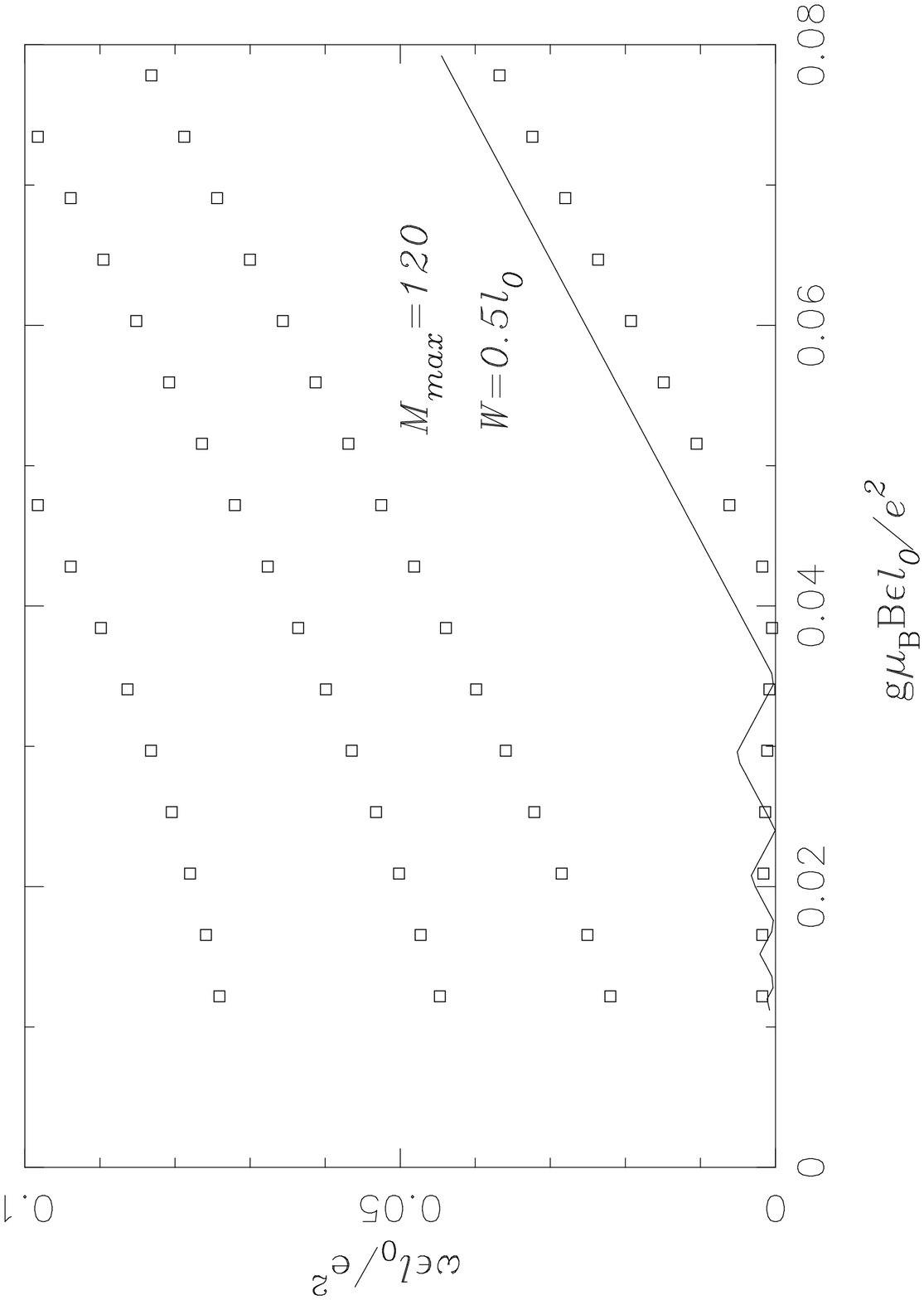}
   }
  \hss}
 }
\caption{
Lowest energy $m=0$ excitations for a system with finite
thickness $W=0.5l_0$,
as a function of (a) inverse system size, and (b)
$\tilde{g}$.  The line in (a) at the left indicates
the Zeeman gap.  A very low energy mode
appears below this gap.  The solid line in
(b) indicates how this mode is expected to
behave when quantization of $K$ is properly accounted for
(see text).
}
\label{size}
\end{figure}

We first present results obtained in the absence of 
an impurity potential.  
In these calculations, we focus for concreteness on the case of quasihole
excitations; identical results however would be obtained
for quasielectron states due to an exact particle-hole
symmetry\cite{hf}.  The full excitation spectrum 
(details will be presented elsewhere\cite{future})
includes spin-wave excitations which are not localized 
near the STQ, localized excitations which correspond to 
transitions between different internal states of the STQ, and 
zero-energy excitations which correspond to transitions between
different angular momenta states and hence to translations of the STQ
with a given internal quantum number.  To identify transitions between
different internal energy states we have picked out the portion 
of the spectrum which is independent of the finite radius of 
the electron disk used for our Hartree-Fock calculations 
as illustrated for a particular value of $\tilde g$ 
in Fig.\ref{size}(a).  Spin-wave states 
have a minimum energy\cite{kallin} equal to $\tilde g$ and 
the number of these states in a given energy interval increases
with increasing system size as the continuous spectrum 
of the thermodynamic limit is approached. 
The excitation energy of the single very low energy mode below the 
the Zeeman gap is independent of system size and we identify this
as a transition between internal states of the STQ. 
An analysis of the residue of this pole shows that the
excitation is one in which the transverse  
(perpendicular to the Zeeman field) component of the 
spin-polarization becomes time-dependent (as expected
from its conjugate\cite{hcm,nayak} relationship to $K$),
although there is also a small
``breathing'' motion in the charge density.
We remark that the existence of these low-energy excitations of 
STQ's is not apparent in the `band' structure
of the broken symmetry HF mean-field ground state. 
They are captured only when the time and space dependence of the 
Hartree and exchange local fields is included in the response 
function calculation.  In a classical calculation these modes would be
gapless but a finite gap is expected\cite{raj} in quantum calculations because
of the microscopic size of the STQ's.  
Similar low-energy modes have been identified
in lattice simulations of easy-plane ferromagnets\cite{wysin}.

Fig.\ref{size}(b) illustrates how the internal excitation 
energy depends on of $\tilde{g}$ for a fixed system size and 
provides additional support for our interpretation of this pole.
It appears in our response functions 
{\it both} for large Zeeman coupling and for small Zeeman coupling where 
the ground state has $K \ne 0$.   The collective mode 
frequency vanishes {\it precisely} at 
$\tilde g = g_c $,  where $g_c$ is the value for
which $K \ne 0$ STQ's first become 
stable in the Hartree-Fock approximation.
As $\tilde g$ decreases  
below $g_c $ we expect a series of level crossings in which STQ's with 
larger internal quantum number become the 
ground state\cite{oaknin,hcm}. 
Although the Hartree-Fock ground state calculations
do not respect the quantization of $K$, 
comparison with exact diagonalization calculations suggests 
that accurate estimates of $U_K$ are obtained by 
imposing the condition $K=\sum_m |u_m|^2=$ integer 
as a constraint on the trial wavefunction
in Eq. (\ref{hfstate}).~\cite{aboulfath}.  The 
energy difference between the ground and lowest excited STQ states
obtained by this procedure is illustrated
as a solid line in Fig. \ref{size}(b).  Although the scale of 
the internal excitation energies found in our response 
function calculations agrees with the scale 
associated with this sequence of level crossings, we do
not appear to see vanishing gaps at level crossing positions.  
We remark in passing that the existence of these low energy 
internal excitations should be important in the behavior of 
the magnetization of a quantum Hall system doped away from 
$\nu=1$~\cite{barrett} as a function of $\tilde g$
at temperatures below the Zeeman gap. 
In particular cusps associated with level crossings between
isolated STQ states with different numbers of reversed spins may survive 
weak STQ-STQ interactions and finite temperatures.

We now turn to an analysis of our results for response functions
in the presence of an impurity potential.  
Specifically we consider the effect of an ionized donor 
set back from the electron layer.  
Effects of such impurities on
electromagnetic absorption
have been observed experimentally\cite{richter}. 
Here we analyze this situation 
taking into account that such impurities may bind textured
quasiparticles, using our TDHFA method.

\begin{figure}
 \vbox to 6.0cm {\vss\hbox to 8cm
 {\hss\
   {\includegraphics{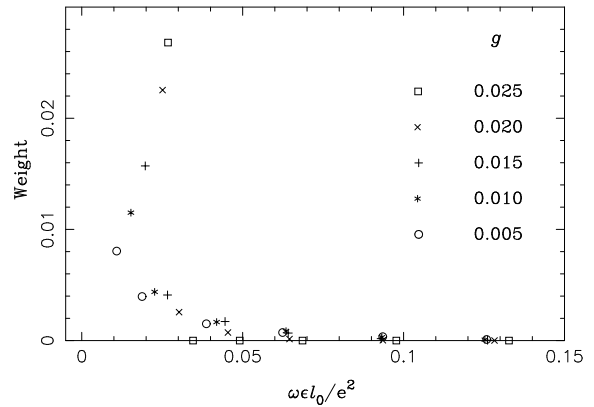}
   }
  \hss}
 }
\caption{Electromagnetic absorption spectrum computed
using the TDHFA for various values of $\tilde{g}$.  Each
symbol represents a delta-function peak with relative
weights for power absorption given by the abscisca.
For $\tilde{g}=0.025$ the spin-texture has collapsed,
and there is only one delta-function peak in the
absorption spectrum.  Lower values of $\tilde{g}$
all have non-trivial spin-textures.  Data shown is
for $M_{max}=120$ and a finite well width of 
$0.5 l_0$ is included in the calculation.  An negatively
charged
impurity is located $2.0l_0$ from the electron plane.
}
\label{abs}
\end{figure}

In the absence of a spin-texture, the response of a
$K=0$ quasiparticle bound to an impurity
to a time-dependent electric field is easy to understand.
The quasiparticle can absorb one quantum of orbital 
angular momentum,
and a single sharp
line is found in the absorption spectrum.
The frequency of this line is independent of the strength
of the Zeeman coupling, $\tilde{g}$.  Physically,
this absorption line corresponds to exciting the
quasiparticle into an orbiting state around the
impurity.

The presence of a spin texture makes this situation
far more rich and interesting.  Fig. \ref{abs} illustrates
the absorption spectrum for various values of $\tilde{g}$.
As can be seen, both the magnitude and the peak frequency
of the absorption is dependent on $\tilde{g}$, a direct
consequence of spin-charge coupling intrinsic to spin
textures in the quantum Hall system. 
(A similar sensitivity of the absorption lineshape to $\tilde{g}$
also occurs for excitations to higher Landau levels\cite{future}.)
When $\tilde{g}$ is moderately large (but not so large as
to completely collapse the spin texture) and the
impurity potential is not too strong, there
is a sharp line in the absorption spectrum that may
be interpreted as exciting the STQ into an orbit
around the impurity center.  Generally, the energy
of this excited state is sensitive to the Zeeman
coupling.  

For weaker Zeeman coupling and stronger impurity potentials,
such that the energy of the
orbiting state rises above the
Zeeman gap, the speed limit for the skyrmion
is exceeded.  The resulting dissipation
broadens the absorption peak,
as illustrated in Fig.\ref{abs}.
We have confirmed that the absorption illustrated
here is indeed a broadened peak in the thermodynamic limit, and not a
series of sharp peaks,
by computing the absorption for larger system
sizes (up to $M_{max}=960$).  We find that the
peak height actually {\it decreases} with increasing
system size, and that the
number of states participating in the absorption
increases.  

In summary, we have demonstrated, using a time-dependent 
Hartree-Fock approach, that spin-charge coupling
leads to a new dissipation channel for spin-polarized
quantum Hall systems. 
The effect is a result of the spin texture
of the quasiparticles, which allows for
low-energy internal modes.  The effect is shown explicitly in
the electromagnetic absorption by quasiparticles
bound to impurities. 

This work was supported in part by 
NATO Collaborative Research Grant No. 930684, by the NSF
under grants DMR-9416906, and DMR-9503814, and by
CICyT of Spain under contract No. MAT 94-0982. 
HAF acknowledges the support of the A.P. Sloan Foundation
and a Cottrell Scholar Award of Research Corporation.


\begin{references}
\bibitem{c1} We note that the matrix element connecting
the $K$ and $K-1$ STQ states with different momenta actually
vanishes in a strictly constant electric field.  However 
variation in the electric field, required to produce 
closed equipotential contours like those on which the network
model is based, will yield a matrix element which permits 
STQ's to radiate spin waves.   We demonstrate this explicitly below for a
skyrmion bound to an impurity.
\bibitem{c2} For typical samples $v_{max} \sim 10^{6} 
{\rm cm/sec}$, which exceeds semiconductor sound velocities,  
but it will be smaller in systems with weaker Zeeman coupling.
Further work will be necessary to determine the relative importance
of phonons and spin-wave radiation in various systems.
\end{references}
\end{document}